\documentclass[preprint]{aastex62}

\usepackage{xcolor}
\usepackage{graphicx} 

\usepackage{amsfonts}
\usepackage{amsmath}
\usepackage{bm}
\usepackage{color}
\usepackage{amssymb}
\usepackage{soul}

\def \beq {\begin{equation}}
\def \eeq {\end{equation}}
\def \bea {\begin{eqnarray}}
\def \eea {\end{eqnarray}}
\def \bfig {\begin{figure}}
\def \efig {\end{figure}}

\def \lab {\label}

\def \de {\partial}

\def \fr {\frac}

\renewcommand{\thefootnote}{${}^\dag$}

\shorttitle{Proton-proton collisions in the turbulent solar wind}
\shortauthors{Pezzi et al.}

\begin{document}

\title{Proton-proton collisions in the turbulent solar wind: Hybrid Boltzmann-Maxwell simulations}

\author[0000-0002-7638-1706]{O. Pezzi}
\affiliation{Gran Sasso Science Institute, Viale F. Crispi 7, I-67100 L’Aquila, Italy}
\affiliation{INFN/Laboratori Nazionali del Gran Sasso, Via G. Acitelli 22, I-67100 Assergi (AQ), Italy}
\affiliation{Dipartimento di Fisica, Universit\`a della Calabria, I-87036 Cosenza, Italy}
\author[0000-0003-1059-4853]{D. Perrone}
\affiliation{ASI - Italian Space Agency, via del Politecnico snc, 00133 Rome, Italy}
\affiliation{Department of Physics, Imperial College London, London SW7 2AZ, United Kingdom}
\author[0000-0001-8184-2151]{S. Servidio}
\affiliation{Dipartimento di Fisica, Universit\`a della Calabria, I-87036 Cosenza, Italy}
\author[0000-0002-1296-1971]{F. Valentini}
\affiliation{Dipartimento di Fisica, Universit\`a della Calabria, I-87036 Cosenza, Italy}
\author[0000-0002-5981-7758]{L. Sorriso-Valvo}
\affiliation{Nanotec-CNR, Sede di Rende, I-87036 Rende, Italy}
\affiliation{Departamento de F\'isica, Escuela Polit\'ecnica Nacional, Quito, Ecuador}
\author[0000-0002-7412-1660]{P. Veltri}
\affiliation{Dipartimento di Fisica, Universit\`a della Calabria, I-87036 Cosenza, Italy}

\correspondingauthor{Oreste Pezzi}
\email{oreste.pezzi@gssi.it}

\input epsf 

\begin{abstract}
The mechanism of heating for hot, dilute and turbulent plasmas represents a long-standing problem in space physics, whose implications concern both near-Earth environments and astrophysical systems. n order to explore the possible role of inter-particle collisions, simulations of plasma turbulence --in both collisionless and collisional regime-- have been compared by adopting Eulerian Hybrid Boltzmann-Maxwell simulations, being proton-proton collisions explicitly introduced through the nonlinear Dougherty operator. Although collisions do not significantly influence the statistical characteristics of the turbulence, they dissipate non-thermal features in the proton distribution function and suppress the enstrophy/entropy cascade in the velocity space, damping the spectral transfer towards large Hermite modes. This enstrophy dissipation is particularly effective in regions where the plasma distribution function is strongly distorted, suggesting that collisional effects are enhanced by fine velocity-space structures. A qualitative connection between the turbulent energy cascade in fluids and the enstrophy cascade in plasmas has been established, opening a new path on the understanding of astrophysical plasma turbulence.
\end{abstract}

\keywords{}

\date{\today}

\section{Introduction} 
\label{sect:intro}
Understanding the dynamics of turbulent and weakly-collisional plasmas represents a challenging problem, whose implications affect a rich variety of systems, ranging from astrophysical environments, e.g.  supernovae remnants, inter-galactic medium and astrophysical jets \citep{parizot06,webb18}, to near-Earth environments such as the solar wind and the planetary magnetospheres \citep{zimbardo10,bruno13,chen16}. In these systems, the energy injected at large scales as gradients is transferred to increasingly larger wave-vectors, producing smaller scale fluctuations, as in typical turbulence processes. Plasma vortices and magnetic coherent structures are routinely recovered in space and astrophysical plasma measurements \citep{servidio12,greco12,perrone16,greco16,perrone17,wang19} and observations support the standard picture of intermittent, inhomogeneous features of the turbulent cascade \citep{sorriso99,veltri99,alexandrova08,sahraoui09,perri12,karimabadiPOP13,bruno15,carbone18}. 

When the turbulent cascade reaches these small spatial and temporal scales, the energy associated with fields fluctuations can be transferred to the particle velocity distribution function (VDF) \citep{servidio15}, where it can be eventually dissipated \citep{vaivads16,sorriso19}. It is widely accepted that large-scale turbulence provides the energy that is dissipated at smaller scales~\citep{verma95,vasquez07,sorriso07, marino08, hadid17} by mechanisms that are not fully determined yet. Several different definitions of dissipation, and diverse dissipative processes, have been invoked in recent years. One approach focuses on the energy dissipation associated with particular phenomena, ranging from specific types of fluctuations (such as kinetic Alfv\'en or whistler waves)~\citep{chandran10,salem12, chang15, gary16, vech17,sorriso18,sorriso19}, to the plasma heating associated with magnetic reconnection~\citep{drake09,servidio11,osman11,servidio12,osman12,daeyoon18,wu13,shay18}. 
Alternatively, the electromagnetic work on particles, namely ${\bm j} \cdot {\bm E}$ ($\bm j$ the electric current density, $\bm E$ the electric field) can be introduced as a surrogate of the dissipation \citep{sundkvist07, wan15}. More recently, the role of the pressure-strain interaction in the energy transfer across scales has been investigated detail \citep{yang17,chasapis18,pezzi19b}. A similar approach, the so-called {\it field-particle correlator}, focuses instead on the analysis of the particular signatures of dissipation mechanisms, such as Landau damping \citep{klein16,klein17,chen19}. Finally, a different approach is based on the role of inter-particle collisions, which introduce irreversibility into the system \citep{tenbarge13,navarro16,pezzi16a,pezzi17a}, so that dissipation is meant as entropy growth. This paper explores the validity of the latter approach.

Even if the common perception is that collisions act, on average, at large characteristic time and space scales \citep{spitzer56,kasper08,maruca11,maruca13,tracy16,chhiber16,vafin19}, they may also be non-negligible at small scales, especially when plasma turbulence develops strong gradients in the velocity space \citep{pezzi16a}. These features are commonly observed in the solar wind and in the terrestrial magnetosheath, manifesting as strong temperature anisotropy enhancements, beams of accelerated particles, rings, and velocity-space vortices \citep{marsch06,servidio15, wilder16, lapenta17}. Because of these observational evidences, a turbulent velocity-space enstrophy cascade has been conjectured \citep{schekochihin16,servidio17} and it has been recently observed in the Earth's magnetosheath and in Eulerian hybrid Vlasov simulations \citep{servidio17,cerri18,pezzi18a}.

Recently, the role of collisions has been considered with novel attention \citep{tatsuno09,tenbarge13,escande15,tigik16,navarro16,pezzi16a,pezzi17a}. The reason for this renewed interest is, at least, two-fold. First, collisions are the mechanism which operates the transition from collisionless (Vlasov) to collisional dynamics, since the collisional operators often satisfy the Boltzmann H-theorem for the entropy growth. This aspect is crucial for investigating the small-scale end of the turbulent cascade, where the physical information contained into phase-space structures needs to be degraded by means of irreversible processes. In order to properly describe the plasma heating from a well-posed thermodynamics viewpoint, the introduction of an irreversible mechanism, such as collisions, is then decisive. Second, it has been recently proposed that plasma collisionality may be enhanced by the presence of fine velocity-space structures, since such structures are rapidly smoothed out by collisions \citep{pezzi16a}. The presence of small-scale phase-space perturbations, incessantly produced by plasma turbulence at kinetic scales, makes collisions act on characteristic times that are much smaller than predicted upon a quasi-Maxwellian assumption, this suggesting a local enhancement of collisionality \citep{pezzi16a}. This paves the way to a novel scenario, where the production of finer velocity-space structures occurs, until the characteristic time associated with their development is balanced by the characteristic time associated with the dissipation of such structures. Very recently, a similar behavior has been also observed in solar-wind data for bi-Maxwellian VDFs \citep{vafin19}. This type of dissipation would act as a purely thermodynamic heating, since the free energy contained into the velocity space structures --which could be converted into other forms of ordered energy by means of several collisionless mechanisms, e.g. micro-instabilities \citep{gary05,matteini12,chenEA16,hellinger17}-- is actually destroyed by an irreversible process. The complete description of such scenario is probably beyond the capability of any present {\it in-situ} observations \citep{pezzi19a}. Addressing it via numerical simulations is therefore of fundamental importance.

\citet{pezzi16a} have described the collisionality enhancement by modeling collisions through the fully nonlinear Landau operator \citep{landau36,rosenbluth57} and focusing on a force-free, homogeneous plasma. This latter assumption represents a caveat that allows to model collisions with a ``proper'' operator, such as the fully nonlinear Landau operator, which can be derived from first-principles and holds the H-theorem for the entropy growth. 
However, the computational cost of the Landau operator is nowadays too demanding for performing high-resolution self-consistent simulations [see \citet{pezzi17a} for further details]. Additionally, the choice of the more general Lenard-Balescu operator \citep{lenard60, balescu60} --which, unlike the Landau operator, takes also into account the presence of spatial fluctuations through the plasma dispersion function-- would make any computational approach unaffordable.

In the present work, we get rid of the approximation of force-free, homogeneous plasma by performing {\it ab-initio} collisional, self-consistent, Eulerian Hybrid Boltzmann-Maxwell simulations of a turbulent plasma. This framework has been widely adopted for describing plasma turbulence at proton and sub-proton scales, using the collisionless Vlasov counterpart \citep{servidio15, valentini16, cerri17, groselj17,perrone18}. In order to make self-consistent simulations affordable, it is necessary to reduce the complexity of the collisional operator. To this aim, we model collisions through the nonlinear Dougherty operator \citep{dougherty64, dougherty67a, dougherty67b}. This is an {\it ad-hoc} collisional operator that has been previously adopted in self-consistent Vlasov-Poisson simulations to describe the collisional dissipation of nonlinear electrostatic waves \citep{pezzi15b} and to model inter-particle collisions in non-neutral plasma columns \citep{anderson07a,anderson07b}. 

In this paper, we compare the collisional (Boltzmann) and the collisionless (Vlasov) model of plasma turbulence by means of direct Eulerian numerical simulations. In particular, starting with the same initial conditions, we analyze Hybrid Boltzmann-Maxwell (HBM) and Hybrid Vlasov-Maxwell (HVM) simulations, investigating whether collisions affect the general properties of turbulence and the generation of non-thermal features in the proton VDF. We also discuss the effect of collisions on the velocity-space cascade \citep{servidio17,cerri18,pezzi18a}. Finally, we focus on the implication of taking into account inter-particle collisions in terms of entropy density \citep{parks12, gary18,liang2019}.

\section{Numerical model} \label{sect:model}

The numerical model employed for this study is based on the HVM system of equations \citep{valentini07}. Here we extend the HVM model by retaining the effect of proton-proton collisions. The dimensionless HBM equations, in presence of collisions, are:
\begin{eqnarray}
 &&\frac{\partial f}{\partial t} + {\bm v} \cdot \frac{\partial f}{\partial {\bm x} } +  \left ( {\bm E} + {\bm v} \times {\bm B} \right) 
\cdot \frac{\partial f}{\partial {\bm v}}= C(f,f) 
 \label{eq:HVMvlas} \\
&&\frac{\partial {\bf B}}{\partial t}\!=\! 
-{\bf \nabla}\!\times\!{\bm E}\!=\!{\bf \nabla}\!\times\!\left[ {\bm u}\times{\bm B}-\frac{{\bm j}\times{\bm B}}{n} + \frac{{\bf 
\nabla}P_e}{n} - \eta {\bm j}\right] \, , \label{eq:HVMfar}
\end{eqnarray}
where $f({\bm x}, {\bm v}, t)$ is the proton VDF, ${\bm E}$ and ${\bm B}$ are the electric and magnetic field, respectively, and $C(f,f)$ is the collisional operator. The current density is ${\bm j}={\bf \nabla}\times{\bm B}$, the proton density $n$ and bulk velocity $\bf u$ are computed as the first two moments of $f$, while $P_e$ is the isothermal pressure of the massless fluid electrons; quasi-neutrality is also assumed. Time, velocities and lengths are respectively scaled to the inverse proton cyclotron frequency $\Omega_{cp}^{-1} = m_p c / e B_0$, to the  Alfv\'en speed $c_A = B_0 / \sqrt{4\pi n_{0} m_p}$, and to the proton skin depth $d_p= c_A/ \Omega_{cp}$, $m_p$, $e$, $c$, $B_0$ and $n_{0}$ being the proton mass, the unit charge, the light speed, the background magnetic field and the equilibrium proton density. Electron inertia effects have been neglected in the Ohm's law, while the small resistivity ($\eta=10^{-3}$) is accurately introduced to suppress numerical instabilities. 

Collisions are modeled through the nonlinear Dougherty operator, which preserves total mass, momentum and energy and also satisfies the H-theorem for the growth of the Gibbs-Boltzmann entropy, as the Landau operator \citep{dougherty64, dougherty67a, dougherty67b}. Comparison with the Landau operator by means of numerical simulations showed that the relaxation towards the equilibrium of an initial VDF is qualitatively and quantitatively similar, upon rescaling time by a constant factor in the Dougherty operator case \citep{pezzi15a}. An important difference between the Dougherty and the Landau operator instead concerns their different effects on the scaling of the phase-space enstrophy cascade \citep{pezzi19a}. The adoption of a nonlinear operator also guarantees that collisional characteristic times are better recovered with respect to a linearized operator, where collisional times are artificially increased \citep{pezzi17a}.

The normalized Dougherty operator reads as:
\begin{equation}
C(f,f)  = \nu \frac{n}{T^{3/2}} \frac{\de }{\de v_{j}} \left[ T \frac{\de 
f}{\de v_{j}} + \left( v -u\right)_j f \right] \ ,
 \label{eq:dg}
\end{equation}
where $\nu$ is the normalized collisional frequency:
\begin{equation}
 \nu = \frac{g \ln\Lambda \beta_p^{3/2}}{16\pi \sqrt{2}\xi}
 \label{eq:nu0}
\end{equation}
$\xi=c_A/c$, $\beta_p=2 v_{th,p}^2/c_A^2$ ($v_{th,p}=\sqrt{k_{_B} T_{0,p}/m_p}$, the proton thermal speed) and $\ln\Lambda$ and $g$ are the Coulombian logarithm and the plasma parameter, respectively. 
The collisional frequency $\nu$ is a numerical parameter that evaluates the strength of the collisional operator with respect to the other terms in the Vlasov equation. By considering the typical solar-wind parameters in Eq. (\ref{eq:nu0}), one gets $\nu\sim 10^{-5}$. 
To appreciate the role of collisions without irremediably increasing their computational cost, we used a collisional frequency $\nu$ that is two orders of magnitude larger than in the typical solar-wind conditions. It is worth stressing the different physical role of $\nu$ and $\eta$. In fact, if $\nu$ is related to proton-proton collisions, $\eta$ could be related to proton-electron collisions \citep{spitzer56}. One may hence argue that, in both collisionless and collisional simulations, proton-electron collisions have been somehow considered in a very simplistic way. However, the presence of $\eta$ here is not intended to mimic any particular physical process.

Equations~(\ref{eq:HVMvlas}) and (\ref{eq:HVMfar}) have been integrated in a $2.5D$-$3V$ phase space domain (i.e., the three-dimensional velocity space is fully described, while, in the physical space, the three vector components depend only on $x$ and $y$). Despite the dimensionality approximation, the $2.5D$ physical space is still able to capture the features of several physical processes \citep{karimabadiSSR2013, wan15, servidio15, LiApJL2016, pezzi17b, pezzi17c, franci18}. The size of the double-periodic spatial domain is $L_x=L_y=L=2\pi\times 20 d_p$ and it is discretized with $512$ grid-points in each direction, while the velocity domain is discretized with $71$ grid-points in the range $v_{j}=\left[-5 v_{th,p},5v_{th,p}\right]\;(j=x,y,z)$ with the boundary condition $f(v_j>5 v_{th,p})=0$. The numerical resolution has been chosen to describe two wavenumber decades in the physical space: one above and one below the proton skin depth $d_p$, close to which the usual spectral break is thought to occur \citep{bruno13}. Although this limitation affects both large and small scales \citep{parashar15}, we are confident that ion scales are well represented \citep{valentini14}. A similar argument holds also for velocity space. The velocity-space resolution sets the scale on which the filamentation instability \citep{parker15,pezzi16b}, which produces an artificial entropy growth, occurs. Here, it has been set in such a way to significantly limit the entropy increase in the collisionless case (See Sect. \ref{sect:collheat}). A detailed description of the HVM algorithm can be found in  \citet{valentini07} and \citet{vasconez15}. The basic current-advance-method algorithm, implemented in the collisionless version of the code, is here modified by introducing the collisional step \citep{filbet02,pezzi13}. 

\begin{figure*}[!thb]
\includegraphics[width=\textwidth]{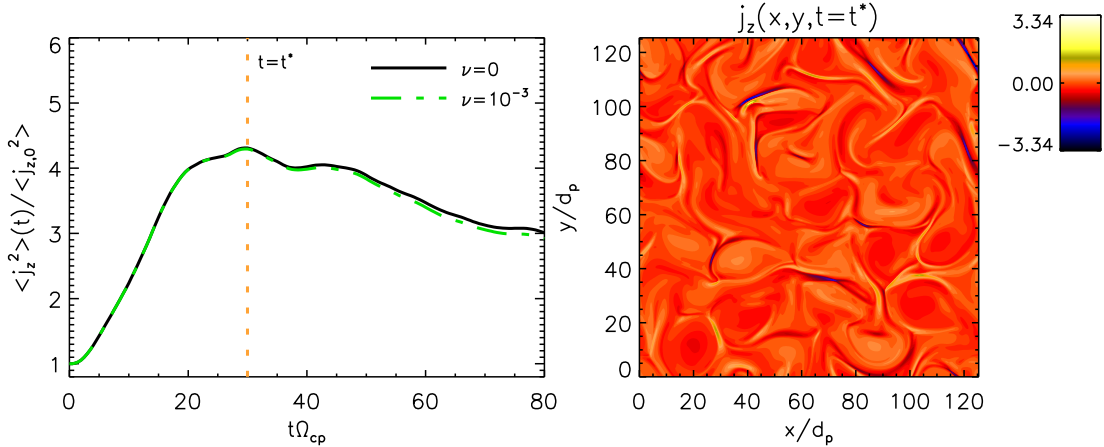}
\caption{(Color online) Left: Temporal evolution of $\langle j_z^2\rangle(t)$ for the Vlasov (black solid line) and Boltzmann (green dash-dotted line) runs. The vertical orange dashed line indicates the peak of the turbulent activity ($t^*=30 \Omega_{cp}^{-1}$). Right: Contour plot of $j_z(x,y)$ for the collisional case at $t=t^*$. The same pattern is found in the collisionless case.}
\label{fig:mjz2}
\end{figure*}

The initial equilibrium is composed of a homogeneous, Maxwellian proton VDF, embedded in an uniform out-of-plane magnetic field ${\bm B_0} = B_0 {\bm e}_z$ ($B_0=1$) and $\beta_p=2$. The equilibrium is then initially perturbed by imposing large-amplitude magnetic field $\bm{\delta b}$ and bulk velocity $\bm{\delta u}$ perturbations. The energy is injected in the wave-number range $k\in[0.1,0.3]$, being $k=m k_0$ with $ 2\leq m \leq 6$ and $k_0=2\pi/L$, in such a way that the spectrum is initially flat; phases are random. The r.m.s. level of fluctuations is $\delta b/B_0 = 1/2$. No density perturbations nor parallel perturbations are introduced at $t=0$. Two different runs have been performed, that differ only by the presence of the collisional operator, while the equilibrium background and the perturbations amplitude are the same. In particular, the first run is collisionless (Vlasov), namely $\nu=0$; while the second run is weakly collisional (Boltzmann), with $\nu=10^{-3}$. We follow the evolution of the system up to $t_{fin} = 80\Omega_{cp}^{-1}$. The computational cost of the two simulations is large, consisting of $\sim1$ million CPU-hours on the supercomputer MARCONI at CINECA, with a massively parallelized and optimized code.

\section{The role of collisions} 
In the present section we analyze the effects of collisions on the plasma dynamics, by focusing in particular on i) the generation of a turbulent scenario at proton scales; ii) the production of non-Maxwellian features in the proton VDF; and iii) the presence of a phase-space enstrophy cascade.

\subsection{Evolution of turbulence at proton scales}
\label{sect:turb}
At the beginning of the simulations, initial perturbations nonlinearly couple and produce a cascade towards smaller scales. The generation of small-scale fluctuations can be appreciated in the temporal evolution of $\langle j_z^2\rangle$ shown in left panel of Fig. \ref{fig:mjz2}, where $\langle \cdots \rangle$ denotes the average on the spatial domain. We observe that $\langle j_z^2\rangle$ initially increases, then saturates at an almost constant value, in the temporal range $t\in[20,40] \Omega_{cp}^{-1}$, and finally decreases for the presence of numerical dissipation induced by the finite mesh-size of the numerical grid and the small value of the numerical resistivity \citep{valentini14}. The peak of the turbulent activity, indicated with a vertical orange dashed line, occurs at $t^* \Omega_{cp} = 30$. Proton collisions do not seem to have a noticeable effect in the evolution of the current density. Indeed, the temporal evolution of $\langle j_z^2\rangle$ is comparable in the Vlasov (black solid line) and Boltzmann (green dash-dotted line) cases.

\begin{figure}[!t]
\centering
\includegraphics[width=7.5cm]{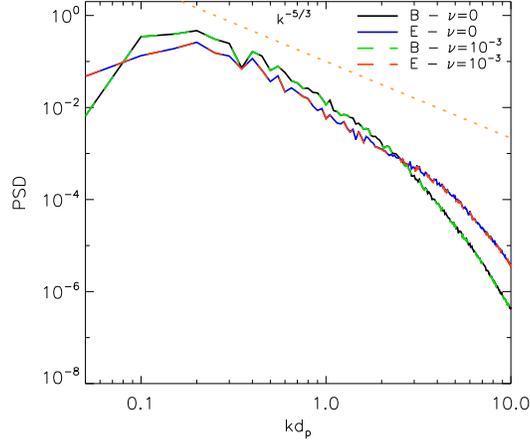}
\caption{(Color online) Omni-directional PSDs of the magnetic field for the collisionless (black solid) and  collisional (green dashed) cases; and of the electric field for the collisionless (blue solid) and collisional (red dashed) cases. The orange dashed line indicates the Kolmogorov expectation.}
\label{fig:spectra}
\end{figure}

The right panel of Fig. \ref{fig:mjz2} shows the contour plot of the out-of-plane current density, $j_z(x,y)$, at $t=t^*$ for the collisional case. This quantity, related to the in-plane magnetic field gradients (the dominant component in our simulation), exhibits a turbulent and intermittent pattern, characterized by the presence of vortices, magnetic islands, current sheets and X-points, suggesting also the presence of magnetic reconnection. A similar behavior is observed for $j_z(x,y)$ in the collisionless case (not shown here).

In order to inspect the turbulence evolution at proton and sub-proton scales, we computed at $t=t^*$ the omni-directional (perpendicular) power spectral densities (PSDs) for both magnetic and electric fields, as a function of $k d_p$. Fig.~\ref{fig:spectra} shows the PSDs for the collisionless (solid line) and collisional (dashed line) runs. The electric and magnetic spectra obtained in the two runs perfectly match, and reveal the typical features observed in solar-wind plasma. Indeed, similarly to previous numerical experiments \citep{perrone13,valentini16,pezzi18a}, an inertial-like range is observed, where the magnetic PSD recalls the Kolmogorov prediction (orange dashed-line) \citep{kolmogorov}, although a proper power-law scaling is not observed due to the limited size of the simulation domain. Around $k d_p \sim 1$, the usual spectral steepening is recovered \citep{leamon}. At sub-proton scales, the electric activity becomes dominant \citep{Bale05}, while density fluctuations are always very low (not shown). The statistical analysis of the standard magnetic fluctuations suggests that intermittency is not affected by collisions.

\subsection{Production of kinetic effects}
\label{sect:kin}
Although collisions do not play a significant role in modifying the statistical characteristics of turbulence, they strongly change the production and the evolution of kinetic, non-thermal features. 

\begin{figure*}[!htb]
\includegraphics[width=\textwidth]{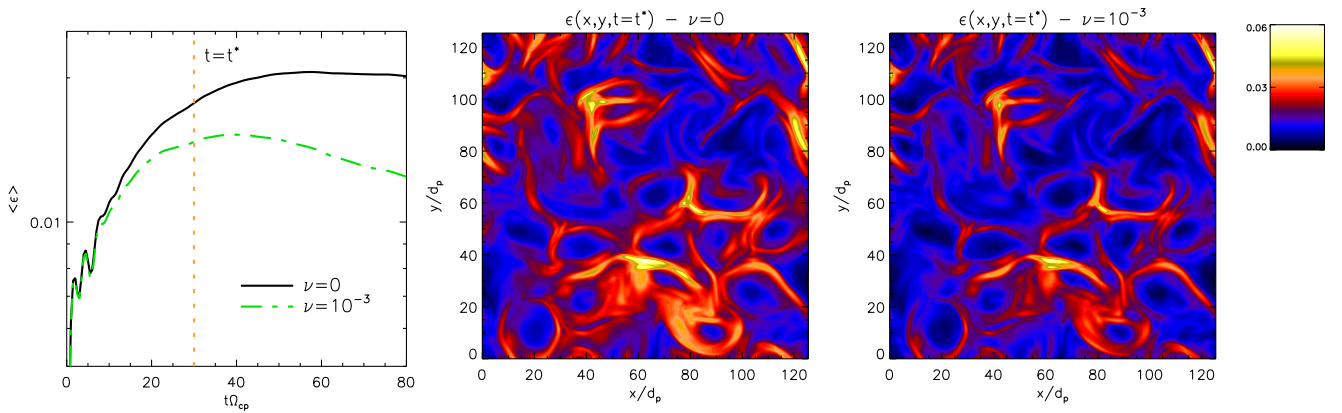}
\caption{(Color online) Left: Temporal evolution of $\langle\epsilon\rangle(t)$ for the collisionless (black solid) and weakly-collisional (green dash-dotted) runs. The orange dashed vertical line indicates $t=t^*=30 \Omega_{cp}^{-1}$. Middle: Contour plot of $\epsilon(x,y)$ at $t=t^*$ for the Vlasov run. Right: Contour plot of $\epsilon(x,y)$ for the Boltzmann simulation.}
\label{fig:eps}
\end{figure*}

In order to quantitatively evaluate the out-of-equilibrium kinetic features in the proton VDF, we make use of the parameter $\epsilon$, defined as follows \citep{greco12,pezzi17b}:
\begin{equation}  
\epsilon({\bm x},t) = \frac{1}{n({\bm x},t)}\sqrt{\int \left[f({\bm x},{\bm v}, t)-g({\bm x},{\bm v},t)\right]^2 d^3v } \, ,
 \label{eq:eps}
\end{equation}
where $g({\bm x},{\bm v},t)$ is the Maxwellian distribution function associated to the observed $f({\bm x},{\bm v},t)$, i.e. with the same density, bulk speed and temperature. The left panel of Figure \ref{fig:eps} shows the temporal evolution of $\langle \epsilon \rangle(t)$. In the collisionless case, $\langle \epsilon \rangle (t)$ increases during the set-up of the nonlinear cascade and then saturates at a nearly constant value. On the other hand, when collisions are in place, $\langle \epsilon \rangle (t)$ slowly decreases after the peak of the turbulent activity (orange dashed line). This global behavior, similar to the enstrophy in fluid flows, suggests that the velocity-space complexity saturates in the ideal (Vlasov) case, while in the Boltzmann plasma there is the tendency to return to Maxwellianity, i.e. $\epsilon \rightarrow 0$. This is consistent with the pattern observed in the contour plots of $\epsilon(x,y,t=t^*)$, where we find a more complex structure in the case without collisions (middle panel), with intense and broad regions of non-Maxwellianity, with respect to the collisional run (right panel). In the latter case, these structures are weaker and narrower. However, in both cases, non-Maxwellian regions are located close to turbulent current sheets.

\begin{figure*}[!htb]
\begin{minipage}{\textwidth}
   \centering
\includegraphics[width=12cm]{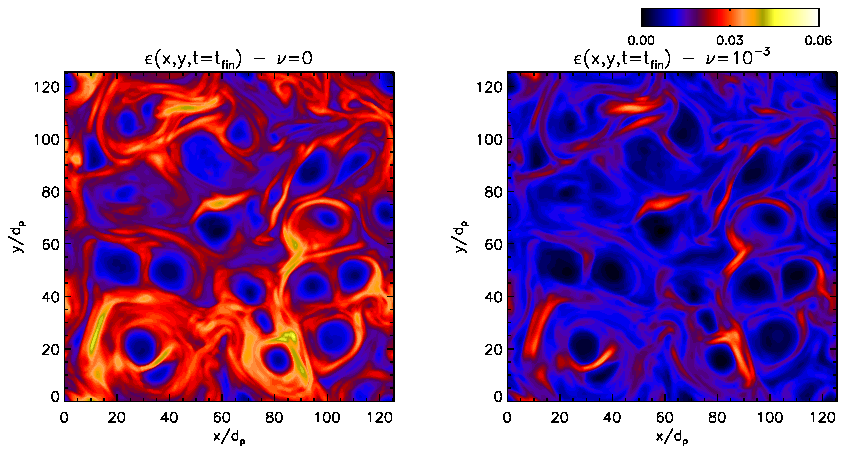}
  \end{minipage}
 \hfill
 \begin{minipage}{\textwidth}\centering
   \includegraphics[width=12cm]{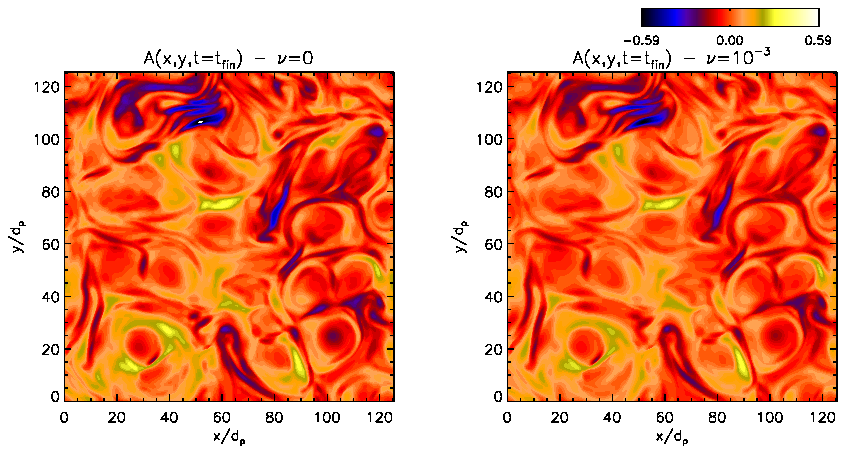}
 \end{minipage}
\caption{(Color online) Contour plots of $\epsilon(x,y)$ (top) and $A(x,y)$ (bottom) at $t=t_{fin}=80\Omega_{cp}^{-1}$ for the Vlasov (left) and Boltzmann (right) cases.}
\label{fig:epsfinaltime}
\end{figure*}

The time evolution of $\epsilon(x,y)$ in the two runs shows substantial differences, as seen by comparing Fig. \ref{fig:eps} with the top panels of Fig. \ref{fig:epsfinaltime}, where maps of $\epsilon(x,y)$ at the final time of each run are shown. It is evident that collisions strongly reduce $\epsilon(x,y)$ with time in the whole volume, effectively leading the plasma towards the thermal equilibrium, except for small regions near the turbulent current structures. On the other hand, in the collisionless case, the areas with large $\epsilon(x,y)$ slightly spread around the current sheets, thus increasing the proton global deviation from Maxwellian with time. Additionally, the amplitude of the most intense values weakly decreases with time. This is probably due to collisionless processes, such as kinetic instabilities, that drive back the free energy contained in non-equilibrium features of the proton VDF into the electromagnetic fields \citep{hellinger17}. 

Bottom panels of Fig. \ref{fig:epsfinaltime} show the proton temperature anisotropy $A = 1 - T_\perp/T_\parallel$, evaluated with respect to the background magnetic field, at the end of each run. No significant differences are found in the two simulations, suggesting that collisions dissipate purely kinetic characteristics --i.e. which cannot be interpreted in terms of anisotropic pressure tensor models \citep{cgl56}-- much faster than temperature anisotropies. This result may be also interpreted as an evidence that collisions smooth fine velocity-space structures on different characteristic times, which depend on the considered velocity scales, as suggested by \citet{pezzi16a} and \citet{pezzi17a}. Finally, note that it can be expected that the long-term evolution of the HBM simulation would converge to the Maxwellian equilibrium, since the Dougherty operator satisfies the H-theorem.

\subsection{Phase-space enstrophy cascade}
\label{sect:Hermite}
To further point out how collisions affect the presence of non-thermal features in the proton VDF, in this section we investigate the development of a phase-space cascade, which is induced by the presence of turbulent fluctuations, as recently proposed in several works \citep{schekochihin16, servidio17,cerri18, pezzi18a, Eyink18}. The idea of this process is that collisionless plasma turbulence initiates the production of a cascade-like process in the full phase-space, leading to the formation of non-Maxwellian features. Here we investigate this phase-space cascade in both Vlasov and Boltzmann simulations.

To analyze the phase-space details, we adopt a $3D$ Hermite transform representation of $f$, namely $f({\bm v}) = \sum_{\bm m} f_{\bm m} \Psi_{\bm m}( {\bm v})$,  where $\Psi_{\bm m}( {\bm v}) = \prod_j \psi_{m_j}( v_j)$ ($j=x,y,z$) and and the 1D basis is:
\beq
\psi_m(v)=\frac{ H_m\!\!\left(\!\fr{v-u}{v_{th}}\!\right) }{\sqrt{2^m m! \sqrt{\pi} v_{th} }} e^{ - \frac{(v-u)^2}{2 v_{th}^2}}.
\lab{eq:psim}
\eeq
In the above mother-function $u$ and $v_{th}$ are the local bulk and thermal speed, respectively; and $m\geq 0$ is an integer (we simplified the notation suppressing the spatial dependence). The eigenfunctions obey the orthogonality condition  $\int_{-\infty}^{\infty} \psi_m(v)\psi_l(v) d v = \delta_{m l}$. Since the basis is opportunely shifted in the local bulk speed frame and normalized to the local density and temperature, we focus on the presence of higher-order fluctuations in the Hermite space. A Gaussian quadrature is also introduced to avoid spurious aliasing and convergence problems \citep{Zhaohua14,servidio17,pezzi18a}. The accuracy of the Hermite transform is finally verified through the Parseval-Plancerel spectral theorem. The Hermite coefficients $f_{\bm m} = \int_{-\infty}^{\infty} f({\bm v}) \Psi_{\bm m}( {\bm v} ) d^3 v$ have been computed for each spatial point. An highly deformed VDF generates plasma enstrophy, defined as
\begin{equation}
\Omega({\bm x}) \equiv \int_{-\infty}^{\infty} \delta f^2({\bm x}, {\bm v}) d^3 v = \sum_{\bm m>0} \left[f_{\bm m}({\bm x})\right]^2.
\label{eq:enst}
\end{equation}
Note that the enstrophy is related to the non-Maxwellian parameter, since $\Omega=\epsilon^2 n^2$, and is also intimately related to the entropy, when the level of fluctuations of the velocity distribution function is small. 

In order to project the VDF over the Hermite basis, we set $N_m=100$ modes in each velocity direction. To reduce the computational effort, the projection has been applied to a subset of the original volume which represents a uniform coarse-graining of the original $512^2$ spatial domain, whose size is $64^2$. The convergence of the Hermite decomposition has been achieved already on a coarser ensemble of $32^2$ VDFs (not shown here).

\begin{figure}[!thb]
\centering
\includegraphics[width=10cm]{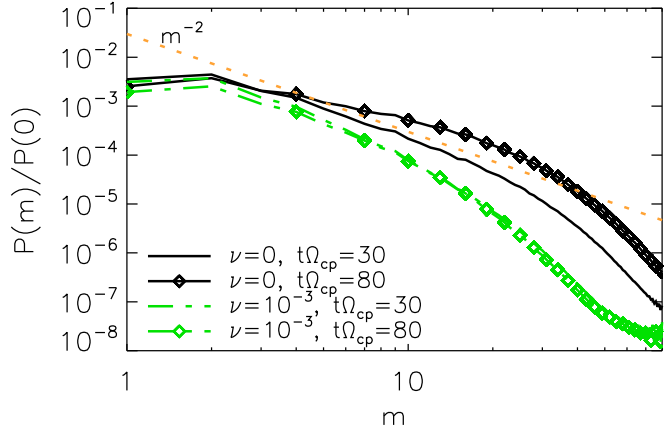}
\caption{(Color online) Isotropic Hermite spectra $P(m)/P(0)$ for the collisionless case at $t=t^*$ (black solid) and $t=t_{fin}$ (black solid with diamonds) and for the collisional case at $t=t^*$ (green dash-dotted) and $t=t_{fin}$ (green dash-dotted with diamonds). The prediction for the magnetized case ($m^{-2}$) is displayed in orange dashed line as a reference. }
\label{fig:hermite}
\end{figure}

From the coefficients $f_{\bm m}({\bm x})\equiv f_{\bm m}({\bm x},{\bm m})$, we computed the enstrophy spectrum $P(m_x, m_y, m_z)= \langle f_{\bm m}({\bm x})^2\rangle$, where $\langle\dots\rangle$ indicates spatial average. The isotropic (omnidirectional) 1D Hermite spectrum is finally obtained by summing $P(m_x,m_y,m_z)$ over concentric shells of unit width, i. e. $P(m) = \sum_{m-\frac12 < |{\bm m'}| \leq  m+\frac12} P({\bm m}')$. Figure \ref{fig:hermite} shows the isotropic Hermite spectra $P(m)/P(0)$ (normalized to the mode $m=0$, which is the only mode excited if the profile is Maxwellian) for the Vlasov and Boltzmann simulations, at both $t=t^*$ and $t=t_{fin}$. As expected, in the collisionless case, a power-law behavior is recovered for a decade of Hermite coefficients \citep{schekochihin08,tatsuno09,kanekar15,parker16,schekochihin16,servidio17,pezzi18a}. The Hermite spectrum breaks around $m\simeq 25$, where the artificial dissipation of the Eulerian scheme may affect the dynamics. At $t=t^*$, the energy distribution is close to the prediction $P(m)\sim m^{-2}$ (orange line in Fig. \ref{fig:hermite}), which corresponds to the case where magnetic fluctuations are dominant in the cascade process \citep{servidio17}. 
At a later stage of the Vlasov simulation, the spectrum becomes shallower, indicating an accumulation of enstrophy at higher Hermite coefficients, still compatible with the slope $m^{-2}$. This accumulation resembles the accumulation of energy in ideal flows, as depicted in the statistical mechanics of complex systems \citep{kraichnan58,frisch95}. The entrophy cascades to finer scales (higher $m$), there accumulates and might flow back to larger $m$ values, similarly to MHD \citep{wan09} and to plasma echo effects \citep{gould67,malmberg68,schekochihin16,parker15,pezzi16b}.

In the collisional case, spectra are less developed, indicating that collisions cancel the finer scale enstrophy. Hence, we here observe the enstrophy dissipation range in plasma turbulence by using a Boltzmann-like simulation: collisions might act as one of the possible mechanisms dissipating the free energy (enstrophy) in the VDF \citep{lesur14,servidio17,pezzi19a}.

This may reveal analogies with the irreversible suppression of the turbulent energy by the standard viscous and resistive dissipation in fluid (Navier-Stokes or MHD) flows, although the two definitions of dissipation (See Sect. \ref{sect:intro}) are in principle different. Fluid dissipative effects irreversibly convert the turbulent energy into heat through viscous-type interactions; while collisions significantly suppress non-Maxwellian features in the proton VDF, that are one of the kinetic-scale counterparts of the fluid turbulent fluctuations \citep{servidio15,klein18}. A second analogy may concern the spatial distribution of dissipative structures. Indeed, the MHD-like dissipation has been found to occur close to enhanced current structures \citep{osman11,osman12,wan15}. Here, we have shown that collisions become effective where the plasma is significantly non-thermal. This can be qualitatively appreciated from the maps of $\epsilon$ and quantitatively from the effects on the Hermite spectra. Then, since the non-Maxwellian structures are co-located with regions of intense current activity \citep{servidio12}, a connection between the effects of collisions and the sites where the MHD-like dissipation of turbulent fluctuations is thought to occur can be suggested. It is worth noting, however, that, with respect the classical dissipation in hydrodynamics, the collisional terms are more non-local and their combined effect in physical and velocity space can be much more complex \citep{pezzi19a}. As can be seen from the Hermite spectra, the action of dissipation is already present at $m\lesssim 10$. This might be due either to the nonlinear structure of the Dougherty operator, or to the value of $\nu$. 

\section{Thermodynamical heating and entropy growth}
\label{sect:collheat}

\begin{figure}[!b]
\centering
\includegraphics[width=9.5cm]{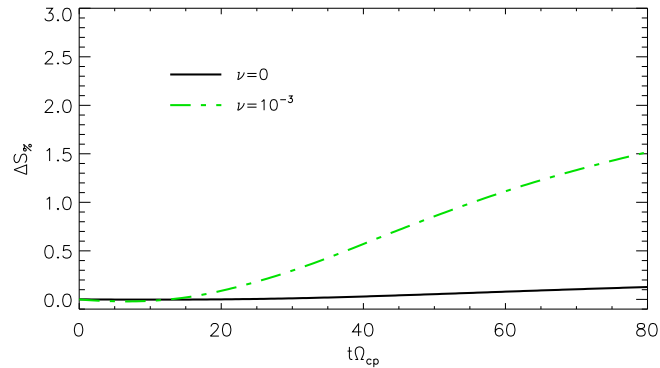}
\caption{(Color online) Temporal evolution of the entropy growth (in percentage with respect to the initial value) for the collisionless (black solid) and  collisional (green dashed) cases.}
\label{fig:entr}
\end{figure}

This last section is dedicated to discuss the effects of collisions in terms of plasma thermodynamics. The spatial average of the proton kinetic temperature, defined as the VDF second-order moment, grows similarly for the two cases (not shown). Indeed, the Dougherty operator, as the Landau one, does not affect the evolution of the second order moment of the proton VDF when considering only proton-proton collisions. Note that this characteristic may be different by including also ion-electron collisions, since the energy transfer between species would be allowed. Conversely, the entropy evolution, being the entropy defined as:
\begin{equation}\label{eq:entr}
 S ( t) = - \int d^3r d^3v f \log{f} \; ,
\end{equation}
shows significant differences in the two cases. Figure \ref{fig:entr} displays the temporal evolution of the entropy growth in the collisionless and weakly-collisional cases, respectively. In the Boltzmann case the entropy increases as a direct effect of the presence of collisions. The very small increase of entropy ($\%0.1$) in the collisionless case (about 1 order of magnitude smaller with respect to the collisional case) is due to the presence of filamentation instability at the velocity-space grid level \citep{pezzi16b}. 

The entropy growth represents a key effect of the introduction of collisions. The VDF free energy, contained in the non-thermal features and available in general for being converted into other forms of ordered energy (e.g. instabilities), is dissipated by collisions and the information contained in such structures, is irreversibly degraded. This ultimately increases the entropy. Although occurring without any variation of the proton temperature, this process still represents a dissipative process, since the system is slowly and irreversibly reaching the thermodynamical equilibrium under the effect of collisions.

We conclude this section by focusing on the entropy density, defined as:
\begin{equation}\label{eq:entrdens}
 s ({\bm x}, t) = - \int d^3v f \log{f} \; .
\end{equation}
This quantity has been widely adopted for describing entropy production in shock waves \citep{parks12} and in collisionless plasma turbulence simulations \citep{gary18}. We would remark that the physical meaning of the entropy density is not related to the Boltzmann thermodynamic entropy $S$ [Eq. (\ref{eq:entr})]. Indeed, while the former is local in physical space, the latter includes global integration over the whole phase space. Furthermore, only the Boltzmann entropy $S$ satisfies the H-theorem. 

Note also that, for small perturbations of the VDF, i.e. $\delta f=f-f_0$, the variation of $s$ can be expressed as:
\begin{equation}\label{eq:s2}
 \Delta s = s({\bm x}, t) - s ({\bm x}, 0) \simeq - \int d^3v \left[ \delta f (1 + \log{f_0}) + \frac{\delta f^2}{f_0} \right] \, ,
\end{equation}
which is similar to $\epsilon$ (i.e. $\epsilon^2 n^2 \simeq \int d^3v \delta f^2$). Both quantities describe the presence of non-Maxwellian features in the proton VDF. In other words, the entropy density $s$ could be adopted as another proxy for highlighting the presence of non-Maxwellian features in the proton VDFs. 

\begin{figure*}[!b]   
\centering
    \includegraphics[width=15cm]{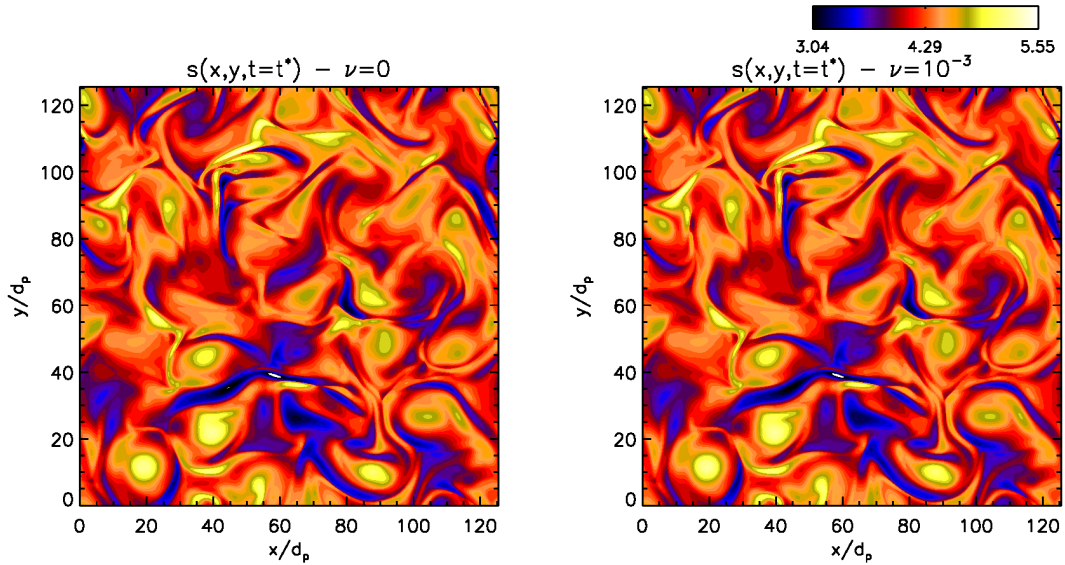}
\caption{ (Color online) Contour plots of $s ({\bm x}, t)$ for both collisionless (left) and weakly-collisional (right) runs at $t=t^*=30\Omega_{cp}^{-1}$.}
\label{fig:entrdens}
\end{figure*}

Figure \ref{fig:entrdens} displays the contour plot of $s(x,y)$ at $t=t^*$, for collisionless (left panel) and collisional (right panel) cases. No significant quantitative nor qualitative differences are found. This similarity is recovered even at different time instants (not shown here). By comparing $s$ (Fig. \ref{fig:entrdens}) and $\epsilon$  (Fig. \ref{fig:eps}), it is evident that the entropy density $s$ also peaks at the center of vortices and magnetic islands, since significant contributions from the pressure terms are expected in these locations. On the other hand, $\epsilon$ peaks in the proximity of the strong current sheets \citep{servidio12,osman11,osman12}, where dissipation of turbulent fluctuations energy is thought to occur. 

Moreover, in the Boltzmann case, $\epsilon$ decreases with time, while $s$ does not show a significant temporal evolution (not shown). This probably indicates that $\epsilon$, unlike $s$, is able to retain the effect of the collisional dissipation. The similar behavior of the entropy density for the HVM and HBM cases may be also explained in terms of Hermite spectra. Indeed, since the Hermite spectra show power-law behavior, we can expect that the summation in Eq. (\ref{eq:entrdens}) is dominated by lower Hermite coefficients, where spectra recovered for the collisionless and collisional cases are rather similar; this implying the similar patterns shown in Fig. \ref{fig:entrdens}. Note that the current definition of entropy density may intrinsically hide differences during the temporal evolution of the system, since it also includes the adiabatic part. As recently proposed by \citet{liang2019}, adopting a velocity-space and a physical-space entropy density may provide further insights.

\section{Conclusions} 
\label{sect:concl}
We have investigated the role of proton-proton collisions, modeled through the Dougherty operator, on the dynamics of weakly-collisional turbulent plasmas by means of direct numerical simulations. 

By comparing the results of collisionless (Vlasov) and collisional (Boltzmann) simulations, we have determined that the statistical properties of plasma turbulence at proton inertial scales are not influenced by inter-particle collisions. On the other hand, the development of kinetic features is strongly suppressed by collisions, which dissipate non-Maxwellian features, driving plasma towards thermal equilibrium. The temporal range analyzed in the simulation is not long enough to allow collisions to effectively lead the system to equilibrium. However, the temporal evolution of the non-Maxwellian parameter $\epsilon$ suggests long-term convergence towards thermalization in the presence of collisions. It cannot be excluded that, at such later stage, collisions may also affect the features of turbulence. 

Although the presence of collisions strongly attenuate the deviations form the thermodynamic equilibrium, the temperature anisotropy is almost not affected. This supports the idea that collisions dissipate different kinetic features on different characteristic time scales. In particular, the dissipation is much faster (i.e. collisionality is locally enhanced) for those phase-space structures that cannot be described in terms of pressure tensor anisotropy \citep{cgl56}, i.e. the ones associated with fine velocity-space structures \citep{pezzi16a}. Collisions act in the phase-space cascade, dissipating enstrophy at the finest scales (thus increasing plasma entropy), similarly to Navier-Stokes turbulence. Similarly to the termination of the cascade in classical fluids, where energy is cancelled by viscous terms at small spatial scales, here we observe that the collisional operator acts at large values of $m$, effectively damping the enstrophy cascade \citep{schekochihin16, Eyink18}. As it can be seen, the rollover of spectra occurs at $m\sim40$ in the collisionless case and at $m\sim20$ for the collisional run. In analogy with the Kolmogorov dissipation scale for hydrodynamic turbulence, a characteristic enstropy-dissipation scale can be defined here, as discussed in \citet{Eyink18}. The intense role of collisions in dissipating non-Maxwellian structures in the proton VDF may be interpreted in terms of inhomogeneous heating \citep{osman11,osman12,servidio12}, since the role of collisions is mainly confined to regions where plasma is non-Maxwellian and these regions are co-located with regions of intense current activity \citep{servidio12}, these being the sites where the MHD-like dissipation of turbulent fluctuations is thought to occur.

Finally, we have analyzed the effect of collisions on the so-called entropy density \citep{parks12,gary18,liang2019}, often adopted to describe the entropy production or to identify eventual sites of dissipation. Since it is not related to the Gibbs-Boltzmann entropy, that satisfies the H-theorem, the first motivation appears questionable. On the other hand, the entropy density may be helpful to identify regions where the plasma is non-Maxwellian. However, at variance with other parameters (here we have considered the $\epsilon$ parameter \citep{greco12}), it has been shown that the entropy density is also peaked inside magnetic islands, where the plasma is close to equilibrium.

Estimating similarities and differences between collisional and collisionless dynamics is of fundamental interest for many complex systems. Dissipation in classical fluids is the transfer of macroscopically organized energy to molecular thermal energy. The comparison between simulations of the ideal Euler equations and the dissipative Navier-Stokes model has been an outstanding challenge in the past decades \citep{MorfEA80, FrischEA08}, known as the global regularity problem for the Navier-Stokes equation, and listed in the Clay-Millennium Prize list problems. Such problem is intimately related to the question as whether real flows may develop singularities at a finite time. In weakly-collisional plasmas, an equivalent problem could be of fundamental relevance for the Boltzmann-Maxwell system, which ideal counterpart is represented by the Vlasov-Maxwell model. The role of collisions in space plasma has been usually interpreted as a secondary effect, due to the small typical collisional frequency $\nu$. However, similarly to the crucial role played by finite, small viscosity in hydrodynamic turbulence, collisions could be fundamentally important in plasma turbulence. Similarly to dissipative terms in fluid flows --that can become locally extraordinarily intense-- the role of collisions is quite fast when the proton VDFs is far from equilibrium and the associated enstrophy dissipation may represent a significant ingredient of the cascade.

Future developments of the present work include the description of fully-kinetic plasmas, where inter-species collisions are also taken into account. This may help introducing a collisional closure in a collisionless plasma. As an example, the role of the pressure-strain interactions, recently proposed for highlighting the energy transfer in collisionless plasmas \citep{yang17,chasapis18,pezzi19b}, may be directly linked to inter-species collisions.

\begin{acknowledgments}
OP sincerely thanks Dr A. Retin\`o and Prof W.H. Matthaeus for the fruitful discussions. OP and SS are partly supported by the International Space Science Institute (ISSI) in the framework of the International Team 405 entitled ``Current Sheets, Turbulence, Structures and Particle Acceleration in the Heliosphere''. 
DP was partially supported by STFC grant ST/N000692/1. 
Numerical simulations discussed here have been performed on the Marconi cluster at CINECA (Italy), within the projects IsC53\_RoC-SWT and IsC63\_RoC-SWTB, and on the Newton cluster at the University of Calabria (Italy). This project (FV, SS) has received funding from the European Union’s Horizon 2020 research and innovation programme under grant agreement No 776262 (AIDA, www.aida-space.eu).
\end{acknowledgments}

\end{document}